\documentclass[manuscript,nonacm,screen]{acmart}

\usepackage{color, colortbl, xcolor}
\usepackage{url}
\usepackage{subcaption}
\usepackage{textcomp}
\usepackage{soul}
\usepackage{multirow}
\usepackage{enumitem}
\usepackage{mathtools}
\usepackage{siunitx}
\usepackage{array}
\usepackage{colortbl}
\usepackage{hhline}

\usepackage{booktabs} 
\usepackage{array}
\usepackage{xcolor}

\usepackage{booktabs}
\usepackage{makecell}
\usepackage[table]{xcolor}
\usepackage{adjustbox}

\usepackage{booktabs}
\usepackage{makecell}
\usepackage[table]{xcolor}
\usepackage{adjustbox}
\usepackage{array}
\usepackage{graphicx}
\usepackage{tabularx}
\usepackage{float}

\definecolor{poslight}{HTML}{FCE4D6}
\definecolor{posmed}{HTML}{F4B183}
\definecolor{posdark}{HTML}{E67E55}
\definecolor{neglight}{HTML}{DDEBF7}
\definecolor{negmed}{HTML}{9DC3E6}
\definecolor{negdark}{HTML}{5B9BD5}
\definecolor{nscell}{HTML}{F2F2F2}

\newcommand{\coefcell}[3]{%
  \cellcolor{#1}%
  \makecell[c]{%
    \rule{0pt}{1.35em}#2\\[-0.05em]
    (#3)\rule[-0.45em]{0pt}{1.15em}%
  }%
}

\newcommand{\sigpos}[2]{\coefcell{posmed}{#1}{#2}}
\newcommand{\sigposlight}[2]{\coefcell{poslight}{#1}{#2}}
\newcommand{\signeg}[2]{\coefcell{negmed}{#1}{#2}}
\newcommand{\signeglight}[2]{\coefcell{neglight}{#1}{#2}}
\newcommand{\nspos}[2]{\coefcell{nscell}{#1}{#2}}
\newcommand{\nsneg}[2]{\coefcell{nscell}{#1}{#2}}

\newcommand*{\rowstyle}[1]{
  \gdef\@rowstyle{#1}%
  \@rowstyle\ignorespaces%
}

\newcolumntype{=}{
  >{\gdef\@rowstyle{}}%
}

\newcolumntype{+}{
  >{\@rowstyle}%
}

\usepackage{arydshln}
\definecolor{LightGray}{gray}{0.97}
\definecolor{linkColor}{RGB}{6,125,233}
\definecolor{green}{rgb}{0.0, 0.65, 0.31}
\definecolor{bleudefrance}{rgb}{0.19, 0.55, 0.91}
\definecolor{ceruleanblue}{rgb}{0.16, 0.32, 0.75}
\definecolor{grey}{HTML}{969696}
\definecolor{violet}{HTML}{756bb1}
\definecolor{dgrey}{HTML}{01665e}
\definecolor{lgrey}{HTML}{5ab4ac}
\definecolor{dgreen}{HTML}{005a32}
\definecolor{purple}{HTML}{ae017e}

\definecolor{editCol}{HTML}{000000}
\definecolor{maskCol}{HTML}{c51b7d}
\definecolor{lrColor}{HTML}{8856a7}
\definecolor{trColor}{HTML}{d01c8b}
\definecolor{ctColor}{HTML}{4dac26}
\definecolor{brickred}{HTML}{f03b20}
\definecolor{improveCol}{HTML}{253494}
\definecolor{worsenCol}{HTML}{d7191c}
\definecolor{DarkBlue}{HTML}{00008B}
\definecolor{mscolor}{HTML}{01665e}
\definecolor{nmscolor}{HTML}{bf812d}
\definecolor{lgreen}{HTML}{ccece6}
\definecolor{dolive}{HTML}{308014}

\definecolor{editCol}{HTML}{000000}
\definecolor{maskCol}{HTML}{c51b7d}
\definecolor{lrColor}{HTML}{8856a7}
\definecolor{trColor}{HTML}{d01c8b}
\definecolor{ctColor}{HTML}{4dac26}
\definecolor{brickred}{HTML}{f03b20}
\definecolor{improveCol}{HTML}{253494}
\definecolor{worsenCol}{HTML}{d7191c}
\definecolor{lgreen}{HTML}{e0f3db}
\definecolor{dpink}{HTML}{CD1076}
\definecolor{pink}{HTML}{FED2D2}
\definecolor{soothinggreen}{HTML}{4dac26}
\definecolor{darkred}{HTML}{8B0000}

\definecolor{dblue}{HTML}{104E8B}
\definecolor{violet}{HTML}{8A2BE2}
\definecolor{mscolor}{HTML}{01665e}
\definecolor{nmscolor}{HTML}{d8b365}
\definecolor{deepgrey}{HTML}{525252}
\definecolor{dslate}{HTML}{2F4F4F}
\definecolor{dolive}{HTML}{556B2F}
\definecolor{teal}{HTML}{388E8E}
\definecolor{mscolor}{HTML}{01665e}
\definecolor{nmscolor}{HTML}{d8b365}

\definecolor{aicolor}{HTML}{018571}
\definecolor{occolor}{HTML}{ff7799}

\definecolor{srcolor}{HTML}{e34a33}
\definecolor{smcolor}{HTML}{253494}
\definecolor{srsmcolor}{HTML}{7fcdbb}
\definecolor{bothcolor}{HTML}{fe9929}
\definecolor{onecolor}{HTML}{018571}
\definecolor{marroon}{HTML}{881c1c}

\colorlet{tablerowcolor4}{gray!50} 

\newcommand*{\textlabel}[2]{%
  \edef\@currentlabel{#1}
  \phantomsection
  #1\label{#2}
}

\colorlet{tableheadcolor}{gray!25} 
\colorlet{tablerowcolor}{gray!15} 
\colorlet{tablerowcolor2}{gray!45} 
\colorlet{tablerowcolor3}{gray!25} 

\newif{\ifhidecomments}
  \hidecommentsfalse 
\ifhidecomments
    \newcommand{\aryan}[1]{}
    \newcommand{\eshwar}[1]{}
    \newcommand{\koustuv}[1]{}
\else
    
    \newcommand{\eshwar}[1]{\textbf{\small\sffamily{\textcolor{soothinggreen}{[#1 -- Eshwar]}}}}
    \newcommand{\koustuv}[1]{\textbf{\small\sffamily{\textcolor{dpink}{[#1 -- Koustuv]}}}}
  \fi

\renewcommand{\textrightarrow}{$\rightarrow$}

\colorlet{tableheadcolor}{gray!25} 
\colorlet{tablerowcolor}{gray!5} 

\definecolor{neutralCol}{HTML}{dd1c77}
\definecolor{neutralGreen}{HTML}{31a354}
\definecolor{NewBlue}{HTML}{1879ba}
\definecolor{bleudefrance}{rgb}{0.19, 0.55, 0.91}  
\definecolor{AfTrColor}{HTML}{0868ac}  
\definecolor{BfTrColor}{HTML}{a8ddb5}  

\definecolor{AfCtColor}{HTML}{b10026}  
\definecolor{BfCtColor}{HTML}{fd8d3c}

\graphicspath{ {figures/} }

\AtBeginDocument{%
  \providecommand\BibTeX{{%
    \normalfont B\kern-0.5em{\scshape i\kern-0.25em b}\kern-0.8em\TeX}}}

\begin{document}


\title[How People Evaluate AI-, Expert-, and Peer-Style Financial Advice]{How People Evaluate AI-, Expert-, and Peer-Style Financial Advice}

\author{Aryan Ramchandra Kapadia}
\orcid{0009-0007-9134-603X}
\email{kapadia8@illinois.edu}
\affiliation{%
  \institution{University of Illinois Urbana-Champaign}
  \city{Urbana}
  \state{IL}
  \country{USA}
}

\author{Eshwar Chandrasekharan}
\orcid{0000-0002-7473-1418}
\email{eshwar@illinois.edu}
\affiliation{%
  \institution{University of Illinois Urbana-Champaign}
  \city{Urbana}
  \state{IL}
  \country{USA}
}

\author{Koustuv Saha}
\orcid{0000-0002-8872-2934}
\email{ksaha2@illinois.edu}
\affiliation{%
  \institution{University of Illinois Urbana-Champaign}
  \city{Urbana}
  \state{IL}
  \country{USA}
}

\renewcommand{\shortauthors}{Kapadia et al.}

\begin{abstract}
As generative AI increasingly becomes a common source of daily decision-making, including financial choices, it is critical to understand how people evaluate AI-generated financial advice.
We conducted a preregistered vignette experiment ($N = 285$) in which substantive financial content---including facts, numerical values, recommendation direction, and core reasoning---was held constant while communication style varied across AI Financial Assistant (AI), Certified Financial Planner (Expert), and Online Community Forum (OC) advice. 
Displayed source attribution was independently manipulated through correctly labeled, unlabeled, and mislabeled conditions, allowing us to separate attribution effects from source-specific communication cues. 
Expert advice was rated more favorably than AI advice on 9 of 10 outcomes ($|d| = 0.20$--$0.47$), and this advantage remained visible without source labels, where Expert advice outperformed AI advice on 8 of 10 outcomes (up to $d = 0.60$). Correct labels added limited differentiation, whereas mislabeling increased ratings of AI advice for \textit{situational fit} and \textit{overall quality} ($d = 0.42$ for each) and attenuated the Expert advantage in \textit{situational fit} ($d = -0.36$). Descriptive analyses further showed that AI advice was most responsive to displayed attribution and, conversely, that advice-style differences were most visible under an AI label. These findings show that financial-advice evaluations are shaped jointly by displayed attribution and message-level communication cues. We position disclosure not as a neutral transparency mechanism, but as an interpretive frame whose accuracy and interaction with message cues can shape trust and reliance.
\end{abstract}



\maketitle


\section{Introduction} \label{section:intro}

The role of generative AI tools is no longer limited to general information-seeking; individuals are increasingly using them to seek guidance in high-stakes decision contexts, including medicine~\cite{ayo-ajibolaCharacterizingAdoptionExperiences2024,shahsavarUserIntentionsUse2023} and finance~\cite{pakHowIndividualsUse2026}. 
In such settings, AI-generated advice is not simply received; it is \textit{evaluated} by users, shaping their trust and reliance~\cite{raeesTrustRelianceMeasurement2026,  raeesPeopleAppropriatelyRely2026,pal2026we}. 
Personal finance is one such consequential domain, where everyday decisions about debt, budgeting, investing, and financial planning can have lasting effects on financial wellbeing~\cite{lusardiEconomicImportanceFinancial2014}. 
People seeking financial guidance encounter AI-generated advice alongside recommendations from professional advisors, friends and family, and online communities. 
Yet it remains unclear how users evaluate this AI-generated financial advice relative to advice from these other sources. 
Understanding these influences is important because users may accept or reject financial guidance not only based on its informational quality, but also on perception cues that may be indirectly related to advice credibility, safety, and trustworthiness.

Prior work presents a mixed picture of how users respond to AI advice. 
In some contexts, people exhibit \textit{algorithm appreciation}~\cite{loggAlgorithmAppreciationPeople2019,schecterAlgorithmicAppreciationAversion2023} and \textit{overreliance}, preferring algorithmic recommendations despite associated errors or costs~\cite{klingbeilTrustRelianceAI2024,rosbachStuckSuggestionsAutomation2026}; in others, people are reluctant to use algorithmic
recommendations, particularly in uncertain decision domains, consistent with \textit{algorithmic aversion}~\cite{dietvorstPeopleRejectAlgorithms2020,mahmudWhatInfluencesAlgorithmic2022}. 
Prior work also indicates that reader preferences can vary with the communication styles of LLM- and human-authored explanations, motivating closer separation of message-level cues from source identity~\cite{zhou2026communication}.
Recent comparisons of LLM and online-community responses have identified differences in communication patterns, including more structured and neutral language in AI responses and greater use of conversational engagement, personalization, and lived experience in community responses~\cite{saha2025ai,saha2026linguistic,yim2026generative}.
More broadly, advice-taking and persuasion research suggests that evaluations may depend on both \textit{source cues}---who the advice appears to come from---and \textit{message cues}---how it communicates reasoning through tone, structure, and framing~\cite{bonaccioAdviceTakingDecisionmaking2006,chaikenHeuristicSystematicInformation1980,hovlandInfluenceSourceCredibility1951,sundarMain2008}. It therefore remains unclear whether differences in how users evaluate AI, expert, and peer advice arise from source attribution, source-specific communication cues, or their interaction.

In this study, we examine how source-specific communication style and displayed source attribution independently and jointly shape evaluations of financial advice. Accordingly, we ask: (1) when substantive financial content is held constant, how do AI-, Expert-, and Online Community-style advice differ in evaluations of presentation and comprehension, safety and risk, source authority, trust, and willingness to follow; and (2) how do displayed source labels shape these evaluations when advice is correctly labeled, unlabeled, or mislabeled? We answer these questions through a preregistered vignette experiment with $N = 285$ U.S.\ adults across eight personal-finance scenarios. We standardize financial facts, numerical values, recommendation direction, and core reasoning across advice styles while independently manipulating displayed source attribution through correctly labeled, unlabeled, and mislabeled conditions.

We find that expert advice was rated more favorably than AI advice on 9 of 10 outcomes ($|d| = 0.20$--$0.47$). These differences were already visible without source labels: in the unlabeled condition, Expert advice outperformed AI advice on 8 of 10 outcomes, with effects as large as $d = 0.60$ for \textit{situational fit}. Correct labels added limited differentiation beyond message-level communication cues, whereas mislabeling selectively increased ratings of AI advice for \textit{situational fit} and \textit{overall quality} ($d = 0.42$ for each). In complementary descriptive sensitivity analyses, AI-style content was most responsive to displayed attribution: presenting it with an Expert label rather than no label improved both \textit{situational fit} and \textit{overall quality} by $d=0.47$. Conversely, differences between the underlying advice styles were most apparent when the advice was displayed with an AI label.

Together, our work contributes a controlled three-source comparison that disentangles source-specific communication cues from displayed attribution, showing that apparent source effects reflect how advice is communicated rather than labels alone. These findings imply that disclosure is not a neutral provenance cue: financial AI interfaces should pair accurate attribution with support for evaluating advice reasoning, assumptions, and risks.
\section{Study Design and Methods}

\begin{figure*}[t]
    \centering
    \includegraphics[
        width=\textwidth,
        trim={0 140 0 35},
        clip
    ]{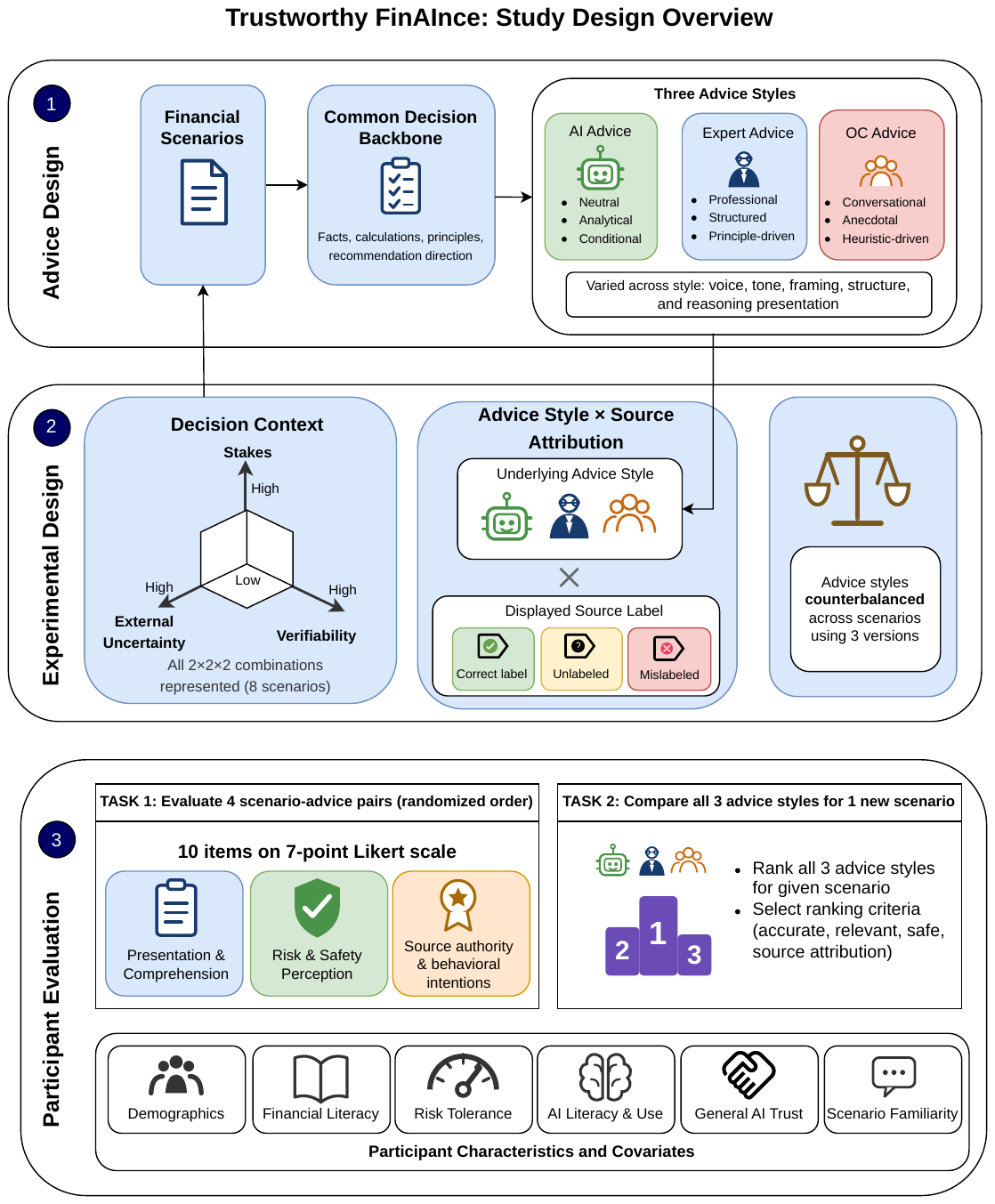}
    \Description{Overview of the study design. A common financial decision backbone is rendered as AI, Expert, and Online Community advice across eight scenarios spanning stakes, external uncertainty, and verifiability. Participants encounter correctly labeled, unlabeled, or mislabeled advice, complete rating and ranking tasks, and provide participant-level moderator measures.}
    \caption{
  Overview of the study design. Eight financial scenarios represented all combinations of stakes, external uncertainty, and verifiability. A common decision backbone was rendered as AI, Expert, and OC (Online community) advice while holding the underlying facts, calculations, financial principles, and recommendation direction constant. Participants evaluated advice under correctly \textit{labeled, unlabeled,} or \textit{mislabeled} source-attribution conditions. They first rated four scenario--advice pairs and then ranked all three advice styles for one additional scenario. Advice styles were counterbalanced across scenarios, and participant-level characteristics and covariates were measured. Note that participants also completed Task 2, but the present analyses focus exclusively on Task 1.
}
    \label{fig:study-design}
\end{figure*}

We conducted a randomized vignette-based survey experiment on Prolific with $N=285$ U.S. adults (see Table~\ref{tab:participant_demographics} for demographics) after excluding incomplete responses, failed attention checks, and submissions completed in under two minutes. The study used eight realistic personal-finance scenarios arranged in a 2×2×2 factorial structure crossing stakes, external uncertainty, and verifiability (Table~\ref{tab:scenario_dimensions}). Participants were assigned to one of two scenario groups and evaluated four scenarios within their assigned group. Within each group, participants were assigned to one of three counterbalanced advice versions. 
Each participant saw all three source styles at least once, with one source style appearing twice. For each scenario, advice versions preserved the same financial facts, numerical values, recommendation direction, and core reasoning, while varying source-specific communication style, tone, and reasoning format: neutral and analytical AI advice, structured and principle-based expert advice, and informal experience-based online community advice.\footnote{For concision, we refer to the three source-specific communication-style conditions as AI, Expert, and OC advice; these terms denote the underlying advice version rather than its displayed source attribution.} Table~\ref{tab:advice_style_operationalization} summarizes the style specifications, and Figure~\ref{fig:annotated_advice_styles} provides matched examples illustrating how the same financial backbone was rendered across conditions. Figure~\ref{fig:study-design} provides an overview of the study design, including advice construction, scenario structure, advice style $\times$ source attribution manipulation, and participant evaluation.

\begin{figure*}[t]
    \centering
    \includegraphics[
         width=0.90\textwidth,
        trim={63bp 568bp 47bp 45bp},
        clip
    ]{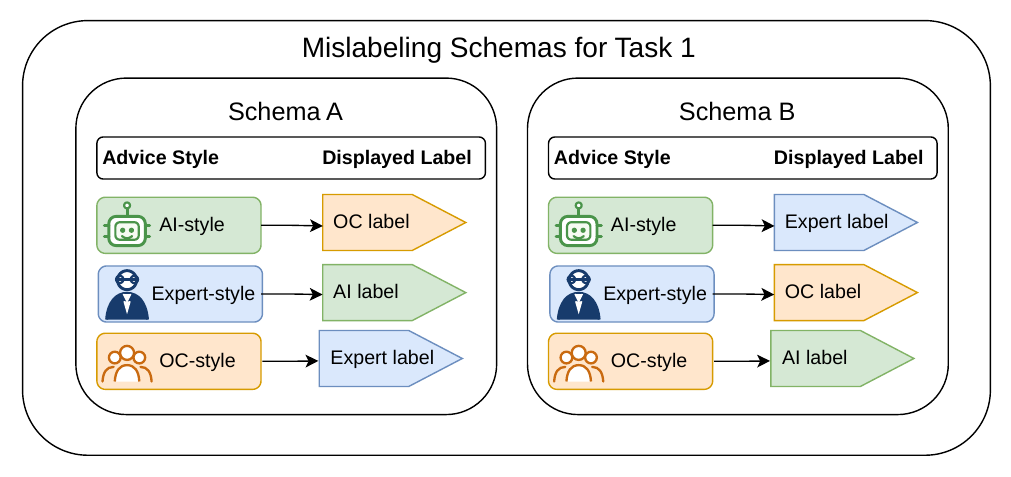}
    \Description{Two diagrams show the counterbalanced mislabeling schemes used in
        Task~1. In Schema~A, AI advice receives an OC label, Expert advice
        receives an AI label, and OC advice receives an Expert label. In
        Schema~B, AI advice receives an Expert label, Expert advice receives
        an OC label, and OC advice receives an AI label.}
    \caption{
  Mislabeling schemas used in Task~1. Participants in the mislabeled
        arm were assigned to one of two counterbalanced label-swap schemas.
        In Schema~A, AI, Expert, and OC advice were displayed with OC, AI,
        and Expert labels, respectively. In Schema~B, they were displayed
        with Expert, OC, and AI labels, respectively. Thus, no advice style
        was paired with its corresponding source label in the \textit{mislabeled}
        condition. Only Task~1 contributes to the analyses reported in this
        paper.
}
    \label{fig:mislabeling_schemas}
\end{figure*}

Participants were randomly assigned to one of three source-labeling arms: \textit{labeled, unlabeled}, or \textit{mislabeled}. The mislabeled arm used one of two counterbalanced label-swap schemas (Figure~\ref{fig:mislabeling_schemas}). After each vignette, participants rated the advice on ten 7-point Likert items capturing \textit{readability, reasoning clarity, situational fit, risk acknowledgment, financial harm risk, misleadingness, perceived source knowledge, overall quality, trust intention,} and \textit{reliance intention} (Table~\ref{tab:measurement_items}). We analyzed responses using mixed-effects regressions with participant random intercepts, scenario fixed effects, and \textit{scenario familiarity} as a covariate (Table~\ref{tab:model_specs}).
The study was approved by the Institutional Review Board at our university. 

\section{Results}

\subsection{Evaluation Differences by Source-Specific Communication Style}

Expert advice was rated more favorably than AI advice on 9 of 10 evaluation dimensions, while online community (OC) advice varied systematically across certain dimensions (Table~\ref{tab:source-hierarchy-mixed}).

\paragraph{Presentation and Comprehension.}
Both Expert and OC advice were rated as more readable than AI advice, with OC showing the largest advantage (Expert vs. AI: $d = 0.47$; OC vs. AI: $d = 0.53$). Reasoning clarity showed a similar pattern: Expert ($d = 0.30$) and OC ($d = 0.17$) outperformed AI, though the effect was stronger for Expert advice. Expert advice was also perceived as a significantly better fit for the protagonist's situation than both AI ($d = 0.39$) and OC ($d = 0.24$).

\paragraph{Risk and Safety Perception}
Expert advice was perceived as safer than AI and OC advice. It was rated as posing a lower risk of financial harm than both AI ($d = -0.27$) and OC advice ($d = -0.23$), while the non-significant AI--OC contrast suggests that participants perceived comparable risk across both non-expert sources. AI advice was rated as more likely to mislead a person with limited financial knowledge than both Expert ($d = -0.30$) and OC ($d = -0.26$). One notable reversal emerged: AI advice was rated as acknowledging risks and uncertainties more than OC advice ($d = -0.17$).

\paragraph{Source authority and Behavioral intentions}
Expert advice presented the clearest advantage on evaluative and behavioral outcomes. Compared with AI, it was rated higher on perceived source knowledge ($d = 0.20$), overall quality ($d = 0.27$), trust intention ($d = 0.24$), and reliance intention ($d = 0.23$). It also outperformed OC across all four dimensions, with small-to-moderate effect sizes ($d$s = $0.19$ to $0.39$). Strikingly, OC was rated significantly \textit{less} knowledgeable than AI ($d = -0.19$) and did not outperform it in overall quality, trust, or reliance.

\subsection{Source Attribution and Advice-Style Interactions}

To evaluate whether displayed source attribution changed advice evaluations beyond the underlying advice styles, we modeled the source-labeling arm using the \textit{unlabeled} condition as the reference (Table~\ref{tab:source-arm-interaction}). In the \textit{unlabeled} arm, participants could already distinguish the three advice styles from message-level communication cues alone. Expert advice was evaluated more favorably than AI advice on 8 of 10 dimensions, including \textit{readability, reasoning clarity, situational fit, financial harm risk, misleadingness, overall quality, trust intention,} and \textit{reliance intention} ($|d| = 0.27$--$0.60$). OC advice was also distinguishable from AI advice on \textit{readability} ($d = 0.43$), \textit{situational fit} ($d = 0.28$), \textit{misleadingness} ($d = -0.27$), and \textit{perceived knowledgeability} ($d = -0.35$), with OC rated as less knowledgeable than AI.

Correct source labels added limited explanatory value beyond these message-level communication cues. The labeled-arm main effect was significant only for \textit{situational fit} ($d = 0.40$), and no Labeled $\times$ Expert or Labeled $\times$ OC interaction reached significance. Mislabeling, however, produced selective shifts in evaluation. Relative to the unlabeled baseline, AI advice shown with a non-AI label received higher ratings for \textit{situational fit} ($d = 0.42$) and \textit{overall quality} ($d = 0.42$). The Mislabeled $\times$ Expert interaction for \textit{situational fit} was also significant ($d = -0.36$), indicating that incorrect labels attenuated the advantage of Expert over AI advice on this dimension. A Mislabeled $\times$ OC interaction ($d = -0.41$) indicated that mislabeling altered the OC--AI difference in \textit{risk acknowledgment}.

\begin{figure*}[t]
    \centering
    \includegraphics[width=\textwidth]{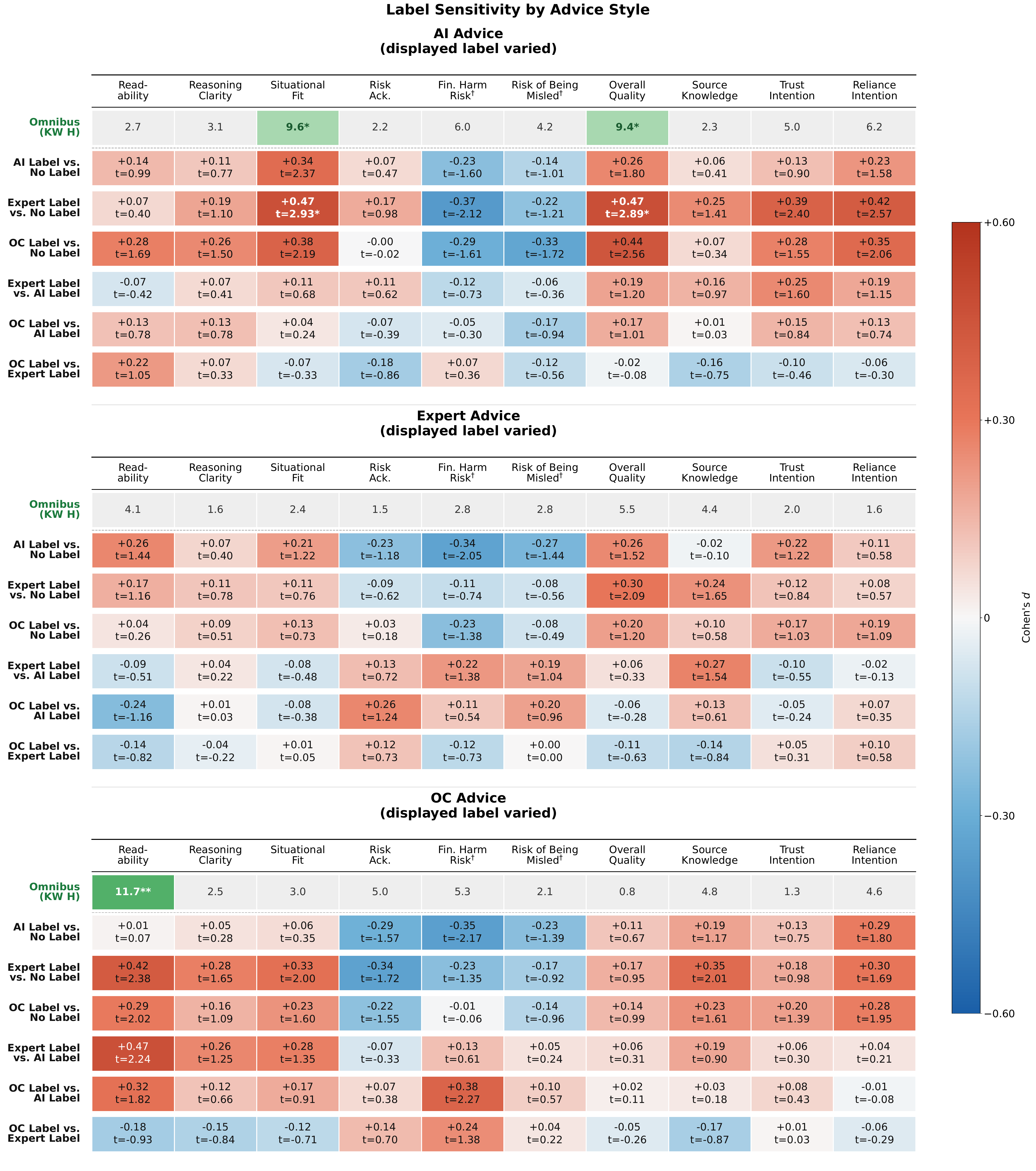}
    \Description{Three heatmap panels show label sensitivity for AI, Expert, and Online Community advice. Each panel holds advice style constant and compares displayed AI, Expert, Online Community, and no-label conditions across ten evaluation outcomes. Cells report Cohen's d and Welch t statistics, with color indicating effect direction and magnitude and an omnibus Kruskal--Wallis column at the left.}
    \caption{
    Label sensitivity by advice style. Each panel holds the underlying advice style constant and varies only the displayed source label. Cell color indicates Cohen's $d$ for each displayed-label comparison, with positive values indicating that the first-named label condition received higher ratings than the second. Cell text reports Cohen's $d$ and the Welch independent-samples $t$ statistic. Pairwise significance stars are based on Holm-adjusted $p$-values within each outcome ($^{*}p < .05$, $^{**}p < .01$, $^{***}p < .001$). The omnibus column reports Kruskal--Wallis $H$ tests across displayed-label conditions. Ratings were first aggregated to the participant $\times$ advice-style $\times$ displayed-label level. Daggered outcomes ($\dagger$) indicate measures for which lower scores are more favorable.
    }
    \label{fig:label-sensitivity}
\end{figure*}

\begin{figure*}[t]
    \centering
    \includegraphics[width=\textwidth]{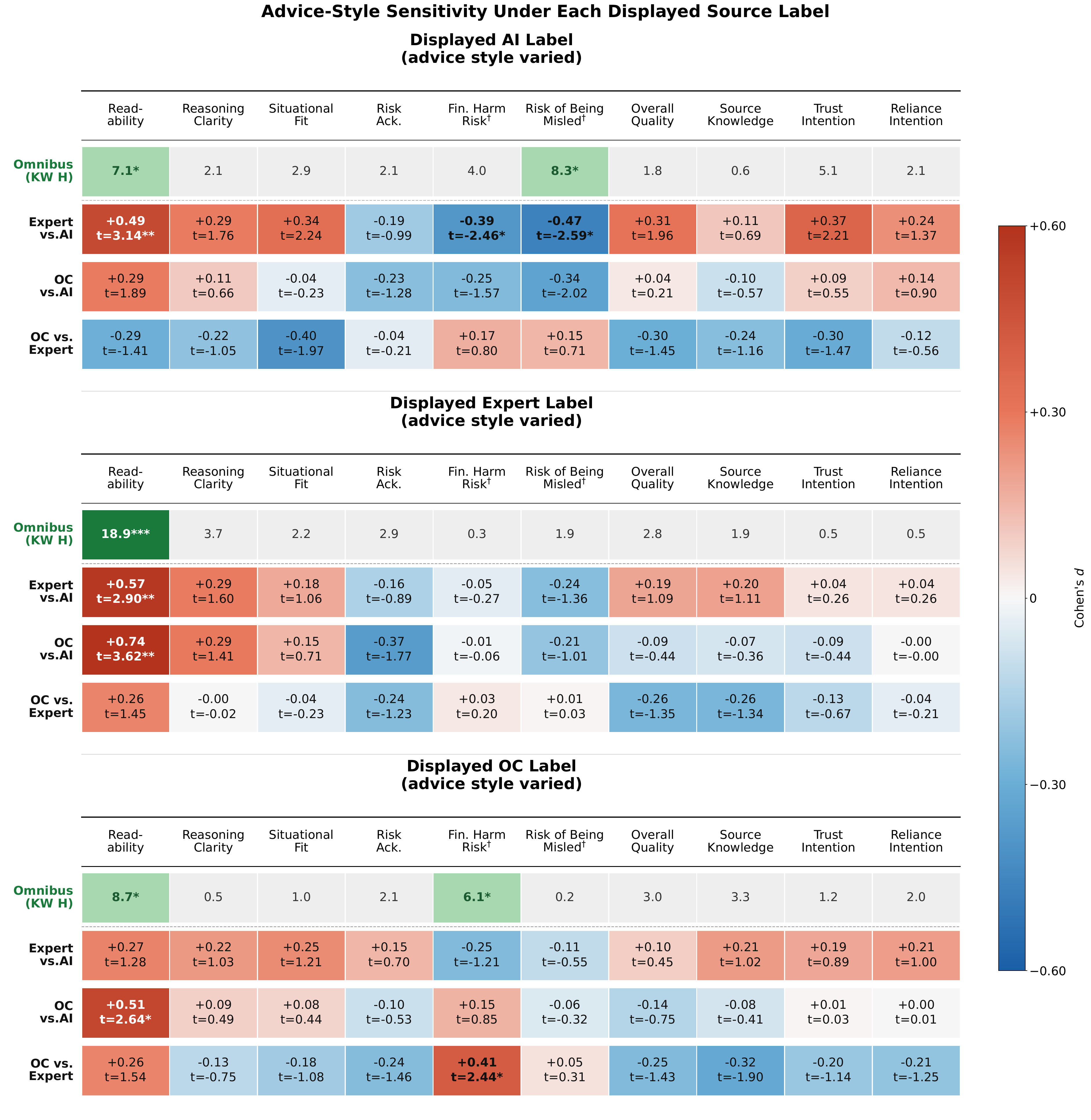}
    \Description{Three heatmap panels show advice-style sensitivity under displayed AI, Expert, and Online Community labels. Each panel holds the displayed label constant and compares underlying AI, Expert, and Online Community advice across ten evaluation outcomes. Cells report Cohen's d and Welch t statistics, with color indicating effect direction and magnitude and an omnibus Kruskal--Wallis column at the left.}
    \caption{
    Advice-style sensitivity under each displayed source label. Each panel holds the displayed source label constant and varies the underlying advice style. Cell color indicates Cohen's $d$ for each advice-style comparison, with positive values indicating that the first-named advice style received higher ratings than the second. Cell text reports Cohen's $d$ and the Welch independent-samples $t$ statistic. Pairwise significance stars are based on Holm-adjusted $p$-values within each outcome ($^{*}p < .05$, $^{**}p < .01$, $^{***}p < .001$). The omnibus column reports Kruskal--Wallis $H$ tests across advice-style conditions. Ratings were first aggregated to the participant $\times$ displayed-label $\times$ advice-style level. Daggered outcomes ($\dagger$) indicate measures for which lower scores are more favorable.
    }
    \label{fig:content-sensitivity}
\end{figure*}

To further visualize the interaction between displayed attribution and message-level communication cues, we conducted descriptive sensitivity analyses on participant-aggregated ratings. In the label-sensitivity analysis, we held the underlying advice style constant and varied the displayed source label (Figure~\ref{fig:label-sensitivity}). Label effects were most pronounced for AI advice: ratings varied across displayed-label conditions for \textit{situational fit} (KW $H = 9.6$) and \textit{overall quality} (KW $H = 9.4$), with the clearest improvements when the same AI advice was presented with an Expert label rather than no label for both \textit{situational fit} ($d = 0.47$) and \textit{overall quality} ($d = 0.47$). Expert advice showed comparatively little sensitivity to relabeling, whereas OC advice showed a more localized label effect for \textit{readability} (KW $H = 11.7$). In the complementary advice-style sensitivity analysis, we held the displayed source label constant and varied the underlying advice style (Figure~\ref{fig:content-sensitivity}). Advice-style differences were most visible under the AI label: Expert advice was rated more favorably than AI advice on \textit{readability} (KW $H = 7.1$, $d = 0.49$) and \textit{misleadingness} (KW $H = 8.3$, $d = -0.47$). Under the Expert label, advice-style sensitivity was concentrated primarily in \textit{readability} (KW $H = 18.9$), with Expert advice ($d = 0.57$) and OC advice ($d = 0.74$) rated as more readable than AI advice. Under the OC label, advice-style differences were again more localized, with OC advice rated as more readable than AI advice (KW $H = 8.7$, $d = 0.51$), but also as posing greater financial harm risk than Expert advice (KW $H = 6.1$, $d = 0.41$).

\section{Discussion}

Our findings show how displayed source attribution and message-level communication cues jointly shape evaluations of AI financial advice, with direct implications for disclosure policy and system design.

\paragraph{The AI evaluation gap is not solely label-driven.}
Expert advice was rated more favorably than AI advice across 9 of 10 dimensions, yet this hierarchy remained identifiable even without any source label: participants distinguished Expert from AI advice based on message-level communication cues. 
This suggests the AI evaluation gap is not purely a labeling artifact; it reflects perceived differences in how AI and human advice communicate. 
This gap is also \textit{dimension-specific}: AI advice was rated as comparable to expert advice on risk acknowledgment, while most of the penalty is concentrated in perceived authority, safety, and source credibility. 
Participants therefore did not simply reject AI advice as insufficiently analytical. 
This aligns with prior work showing that people can evaluate AI-generated support favorably on communicative qualities such as sincerity and actionability, even while recognizing aspects of human support that AI may not replicate~\cite{dasswain2025ai}.
Overall, our findings suggest that AI systems adopting the structured, principle-based communication patterns associated with Expert advice may partially reduce the perceived evaluation gap.

\paragraph{Disclosure effects are limited and accuracy-dependent.}

Correct source labels added limited differentiation beyond what message-level communication cues already conveyed. Participants could distinguish the underlying advice styles even without explicit attribution, suggesting that disclosure alone was not the primary driver of evaluation. However, inaccurate attribution was more consequential: mislabeling selectively weakened distinctions between advice styles that were otherwise visible from message-level communication cues. This asymmetry, in which correct labels offered modest benefits while incorrect labels distorted calibration, suggests that the value of disclosure depends critically on its accuracy. When AI advice is misattributed, inadvertently or by design, inaccurate disclosure may undermine users' ability to calibrate trust to advice quality.

\paragraph{The AI label may heighten scrutiny.}

Differences between advice styles were most visible under the AI label. When advice was labeled as AI, participants distinguished Expert advice from AI advice on \textit{readability} and \textit{perceived risk}, suggesting that AI attribution may heighten attention to message-level communication cues. By contrast, under the Expert label, sensitivity to advice style was concentrated primarily in \textit{readability}, while broader differences in \textit{quality, trust,} and \textit{reliance} were less apparent. This pattern suggests a scrutiny asymmetry: AI labels may induce greater vigilance, whereas Expert labels may encourage greater deference. Source authority, rather than AI identity alone, may therefore shape how carefully people evaluate advice~\cite{chaikenHeuristicSystematicInformation1980, sundarMain2008}.

\paragraph{Design and policy implications.}

Our findings suggest that calibrated evaluation of AI financial advice requires more than source disclosure. Three design directions follow. 
First, interfaces should support advice-level evaluation through explicit reasoning, transparent assumptions, verifiable claims, and calculations, rather than treating source labels as proxies for quality~\cite{huang2026answer}. 
Second, disclosure accuracy matters, not merely disclosure presence. When AI advice is presented as originating from a human expert or online community, users may calibrate their trust against an inaccurate provenance cue rather than the advice's reasoning and quality. 
Third, the scrutiny asymmetry observed in our descriptive analyses suggests that AI labels may encourage more critical evaluation. Rather than minimizing AI attribution to reduce stigma, designers might consider how to preserve its scrutiny-activating properties while also reducing unwarranted discounting of high-quality AI content.

\paragraph{Limitations.}
Our study has several limitations. First, advice style was manipulated as a bundled set of voice, tone, framing, structure, and reasoning features, so we cannot isolate the contribution of any single feature. Second, the stimuli were experimentally constructed and reviewed for information parity; they should not be interpreted as representative of all AI assistants, financial professionals, or online communities. Third, participants evaluated short, text-based vignettes and reported intended trust and reliance rather than making actual financial decisions, limiting behavioral and ecological validity. Fourth, although the scenarios varied in stakes, external uncertainty, and verifiability, the study was not designed to estimate generalizable effects of these dimensions. Finally, the U.S. Prolific sample limits broader generalizability, and the descriptive sensitivity analyses should not be interpreted as confirmatory causal evidence.

Taken together, our results shift the question from whether AI advice should be disclosed to how source attribution interacts with message-level communication cues. In financial contexts, labels are not neutral provenance markers; they frame interpretation and selectively shape judgments of advice quality, trust, and willingness to rely on it.




\bibliographystyle{ACM-Reference-Format}
\bibliography{0paper}


\clearpage

\appendix



\begin{table*}[t]
\sffamily

\centering
\caption{Participant demographics and background characteristics.
Continuous variables are reported as mean (SD); categorical variables
are reported as $n$ (\%).}
\label{tab:participant_demographics}

\begin{tabular}{
    p{0.24\textwidth}
    p{0.45\textwidth}
    p{0.20\textwidth}
}
\toprule
\textbf{Characteristic}
& \textbf{Final analytic sample}
& \textbf{Measure source} \\
\midrule

Sample size
& 285
& --- \\

Age
& 39.76 (SD = 12.22)
& --- \\

Gender
& Male: 143 (50.2\%); Female: 137 (48.1\%);
Prefer not to say: 3 (1.1\%); Missing: 2 (0.7\%).
& --- \\

Household income
& Median category: \$50,000--\$74,999.
& --- \\

Financial literacy (0--3)
& 2.66 (SD = 0.68)
& "Big Three"~\cite{lusardiFinancialLiteracyWorld2011, lusardiEconomicImportanceFinancial2014} \\

Self-rated financial literacy
& Very low: 2 (0.7\%); Low: 20 (7.0\%);
Moderate: 154 (54.0\%); High: 90 (31.6\%);
Very high: 19 (6.7\%).
& --- \\

Financial risk tolerance (1--4)
& No financial risks [1]: 36 (12.6\%);
Average risks [2]: 161 (56.5\%);
Above-average risks [3]: 77 (27.0\%);
Substantial risks [4]: 11 (3.9\%).
& SCF item~\cite{grableAssessingConcurrentValidity2001} \\

AI use frequency
& Never: 9 (3.2\%); Rarely: 20 (7.0\%);
Occasionally: 42 (14.7\%);
Regularly: 80 (28.1\%);
Frequently: 134 (47.0\%).
& --- \\

AI use for personal finance
& No, and have not considered it: 53 (18.6\%);
No, but have considered it: 43 (15.1\%);
Yes, once or twice: 102 (35.8\%);
Yes, multiple times: 87 (30.5\%).
& --- \\

Used AI for finance at least once
& 189 (66.3\%)
& --- \\

General trust in AI (1--7)
& 4.63 (SD = 1.50)
& S-TIAS~\cite{mcgrathMeasuringTrustArtificial2025} \\

Perceived AI literacy (1--7)
& 5.16 (SD = 0.99)
& PAILQ-6~\cite{grassiniPsychometricValidationPAILQ62024} \\

\bottomrule
\end{tabular}
\end{table*}

\begin{table}[t]
\centering
\sffamily

\caption{Scenario-level dimensions in the 2 $\times$ 2 $\times$ 2 vignette design.}
\label{tab:scenario_dimensions}
\begin{tabular}{p{0.10\linewidth} p{0.30\linewidth} p{0.23\linewidth} p{0.23\linewidth}}
\textbf{Dimension} & \textbf{Definition} & \textbf{Low level} & \textbf{High level} \\
\toprule
Stakes &
Magnitude and long-term financial consequences of the decision. &
Smaller or more reversible decisions, such as travel insurance or short-term spending choices. &
Substantial or difficult-to-reverse decisions, such as graduate school, debt repayment, or concentrated investment risk. \\

External uncertainty &
Whether outcomes depend on unpredictable future conditions. &
Outcomes are relatively stable or calculable, such as fixed interest rates or known rent differences. &
Outcomes depend on external events, such as market volatility, weather disruptions, or future job-market conditions. \\

Verifiability &
Whether advice quality can be evaluated against established financial principles. &
Preference-sensitive decisions with no single dominant rule, such as housing lifestyle trade-offs or graduate-school ROI. &
Decisions with broad expert consensus, such as paying high-interest debt or diversifying concentrated assets. \\
\bottomrule
\end{tabular}
\end{table}


\begin{table*}[t]
\centering
\caption{Operationalization of source-specific communication styles.}
\label{tab:advice_style_operationalization}

\setlength{\tabcolsep}{4pt}
\renewcommand{\arraystretch}{1.18}

\begin{tabularx}{\textwidth}{
    >{\raggedright\arraybackslash}p{0.12\textwidth}
    >{\raggedright\arraybackslash}X
    >{\raggedright\arraybackslash}X
    >{\raggedright\arraybackslash}X
}
\toprule
\textbf{Feature}
& \textbf{AI advice}
& \textbf{Expert advice}
& \textbf{OC advice} \\
\midrule

\textbf{Voice}
& Impersonal; no first-person identity
& Mild first-person professional authority
& First-person peer perspective \\

\textbf{Tone}
& Neutral and analytical
& Professional and measured
& Informal and conversational \\

\textbf{Reasoning}
& Explicit analysis and conditional if--then reasoning
& Structured, principle-based reasoning oriented toward long-term planning
& Experiential reasoning supported by practical heuristics \\

\textbf{Framing}
& Balances available options and evaluates financial risk exposure
& Connects the decision to financial foundations and broader goals
& Uses lived experience, subjective judgment, and relatable consequences \\

\textbf{Structure}
& Situation and trade-off; analysis; conditional recommendation
& Quantified framing; trade-off explanation; actionable principle
& Anecdote or position; experiential lesson; direct recommendation \\

\textbf{Excluded by design}
& Personal anecdotes, emotional validation, and unsupported assumptions
& Personal anecdotes, slang, and emotional storytelling
& Formal advisory authority and technical optimization language \\

\bottomrule
\end{tabularx}

\vspace{3pt}

\begin{minipage}{\textwidth}
\footnotesize
\textit{Note.}
We define \textit{advice style} as a controlled bundle of message-level
features involving voice, tone, framing, discourse structure, and reasoning
presentation. Style was varied separately from substantive financial
content: within each scenario, the financial facts, numerical values,
relevant principles, recommendation direction, and core reasoning were held
constant across versions. The three versions were constructed using
source-specific drafting and prompting instructions and were manually
reviewed for information parity and comparable length. The excluded features
represent experimental constraints rather than claims that real-world
sources never exhibit these characteristics. The participant-facing labels were ``AI Financial Assistant,'' ``Certified Financial Planner,'' and ``Online Community Forum.''
\end{minipage}

\end{table*}

\begin{table*}[t]
\centering
\caption{Measurement items used to evaluate financial advice.}
\label{tab:measurement_items}
\renewcommand{\arraystretch}{1.18}
\setlength{\tabcolsep}{6pt}

\begin{tabularx}{\textwidth}{
    >{\centering\arraybackslash}p{0.05\textwidth}
    >{\raggedright\arraybackslash}X
    >{\raggedright\arraybackslash}p{0.25\textwidth}
}
\toprule
\textbf{No.} & \textbf{Measurement Item (7-point Likert scale)} & \textbf{Dimension} \\
\midrule

\multicolumn{3}{l}{\textit{\textbf{Presentation and Comprehension}}} \\
\addlinespace[2pt]
1 & The advice is easy to read and follow.
  & Readability \\

2 & The advice provides clear reasoning for its recommendation.
  & Reasoning clarity \\

3 & The advice fits well with X's situation.
  & Situational fit \\

\addlinespace[4pt]
\multicolumn{3}{l}{\textit{\textbf{Risk and Safety Perception}}} \\
\addlinespace[2pt]
4 & The advice acknowledges potential risks and uncertainties.
  & Risk acknowledgment \\

5 & Following this advice could put X at financial risk.
  & Financial harm risk$\dagger$ \\

6 & Someone with limited financial knowledge could misunderstand this advice.
  & Misleadingness$\dagger$ \\

\addlinespace[4pt]
\multicolumn{3}{l}{\textit{\textbf{Source Authority and Behavioral Intentions}}} \\
\addlinespace[2pt]
7 & The source of this advice appears knowledgeable about financial decisions.
  & Perceived source knowledge \\

8 & Overall, this is high-quality financial advice for X.
  & Perceived overall quality \\

9 & I would trust this advice if I were in X's situation.
  & Trust intention \\

10 & If I were in X's situation, I would feel comfortable following this advice.
  & Reliance intention \\

\bottomrule
\end{tabularx}

\vspace{3pt}
\begin{minipage}{\textwidth}
\footnotesize
\textit{Note.} X denotes the protagonist named in each financial scenario.
Items marked with dagger ($\dagger$) are negatively valenced: higher ratings indicate
greater perceived financial harm or misleadingness.
\end{minipage}
\end{table*}


\begin{table*}[t]
\centering
\sffamily

\caption{Model specifications used in the analysis.}
\label{tab:model_specs}

\begin{tabular}{p{0.30\linewidth} p{0.68\linewidth}}
\toprule
\textbf{Analysis} & \textbf{Model specification} \\
\midrule

Overall advice-style differences &
$Y \sim C(\text{advice\_style})
+ C(\text{scenario})
+ \text{familiarity}
+ (1 \mid \text{participant})$ \\

Advice style $\times$ label arm &
$Y \sim C(\text{advice\_style})
\times C(\text{label\_arm})
+ C(\text{scenario})
+ \text{familiarity}
+ (1 \mid \text{participant})$ \\

\bottomrule
\end{tabular}

\vspace{0.5em}
\begin{minipage}{0.98\textwidth}
\footnotesize
\textit{Note.}
$Y$ denotes each advice-evaluation outcome. All models include
participant-level random intercepts to account for repeated ratings
from the same participant. Scenario fixed effects control for
differences across financial-advice scenarios, and familiarity denotes
participants' self-reported familiarity with the scenario. The
\textit{overall advice-style model} estimates average differences among
AI, Expert, and OC advice across labeling conditions; corresponding
results are reported in Table~\ref{tab:source-hierarchy-mixed}. The
\textit{advice style $\times$ label arm model} estimates whether these
differences vary across the unlabeled, labeled, and mislabeled
source-attribution conditions; corresponding results are reported in
Table~\ref{tab:source-arm-interaction}.
\end{minipage}
\end{table*}

\begin{table*}[t]
\centering
\sffamily

\caption{Overall differences by advice style across advice-evaluation outcomes.}
\label{tab:source-hierarchy-mixed}
\setlength{\tabcolsep}{4pt}
\renewcommand{\arraystretch}{1.35}

\begin{adjustbox}{width=\textwidth}
\begin{tabular}{lcccccccccc}
\toprule
& \multicolumn{3}{c}{Presentation \& Comprehension}
& \multicolumn{3}{c}{Risk \& Safety Perception}
& \multicolumn{4}{c}{Source Authority \& Behavioral Intentions} \\
\cmidrule(lr){2-4}
\cmidrule(lr){5-7}
\cmidrule(lr){8-11}
 &
\makecell{Read-\\ability} &
\makecell{Reasoning\\Clarity} &
\makecell{Situational\\Fit} &
\makecell{Risk\\Ack.} &
\makecell{Fin. Harm\\Risk$^\dagger$} &
\makecell{Risk of Being\\Misled$^\dagger$} &
\makecell{Overall\\Quality} &
\makecell{Source\\Knowledge} &
\makecell{Trust\\Intention} &
\makecell{Reliance\\Intention} \\
\midrule

Expert vs. AI
& \sigpos{+0.410***}{0.47}
& \sigpos{+0.255***}{0.30}
& \sigpos{+0.322***}{0.39}
& \nsneg{-0.107}{-0.10}
& \signeg{-0.334***}{-0.27}
& \signeg{-0.403***}{-0.30}
& \sigpos{+0.263***}{0.27}
& \sigposlight{+0.182**}{0.20}
& \sigposlight{+0.238**}{0.24}
& \sigposlight{+0.233**}{0.23} \\[0.35em]

OC vs. AI
& \sigpos{+0.468***}{0.53}
& \sigposlight{+0.141*}{0.17}
& \nspos{+0.121}{0.15}
& \signeg{-0.183*}{-0.17}
& \nsneg{-0.049}{-0.04}
& \signeg{-0.343***}{-0.26}
& \nspos{+0.016}{0.02}
& \signeg{-0.173*}{-0.19}
& \nspos{+0.048}{0.05}
& \nspos{+0.067}{0.07} \\[0.35em]

Expert vs. OC
& \nsneg{-0.059}{-0.07}
& \nspos{+0.114}{0.13}
& \sigposlight{+0.201**}{0.24}
& \nspos{+0.076}{0.07}
& \signeg{-0.285**}{-0.23}
& \nsneg{-0.060}{-0.04}
& \sigpos{+0.248***}{0.26}
& \sigpos{+0.355***}{0.39}
& \sigposlight{+0.190**}{0.19}
& \sigposlight{+0.166*}{0.17} \\[0.25em]

\midrule
$R^2_m$
& 0.207 & 0.154 & 0.113 & 0.097 & 0.101 & 0.087 & 0.105 & 0.068 & 0.124 & 0.155 \\[0.15em]

$R^2_c$
& 0.399 & 0.374 & 0.381 & 0.352 & 0.460 & 0.405 & 0.389 & 0.314 & 0.369 & 0.376 \\

\bottomrule
\end{tabular}
\end{adjustbox}

\vspace{0.5em}
\begin{minipage}{0.98\textwidth}
\footnotesize
\textit{Note.} This table reports unstandardized mixed-effects regression coefficients $\beta$ with standardized effect sizes $d$ in parentheses. Models estimate overall differences by advice style, averaged across labeling conditions, and include scenario fixed effects, scenario familiarity as a covariate, and participant-level random intercepts. AI is the reference category for Expert vs. AI and OC vs. AI; Expert vs. OC contrasts were computed from the fitted model covariance matrix.
$N=285$ participants and $1{,}140$ vignette-level observations. $R^2_m$ and $R^2_c$ denote marginal and conditional Nakagawa-style $R^2$, respectively.
$^\dagger$ Lower scores indicate more favorable evaluations for these outcomes. Positive coefficients indicate that the first-named advice style was rated higher than the second-named style; for daggered outcomes, negative coefficients indicate more favorable ratings for the first-named source.
$^{*}p<.05$, $^{**}p<.01$, $^{***}p<.001$.
\end{minipage}
\end{table*}

\begin{table*}[t]
\centering
\caption{Advice style and label-arm effects on advice evaluations.}
\label{tab:source-arm-interaction}

\setlength{\tabcolsep}{3.2pt}
\renewcommand{\arraystretch}{1.28}

\begin{adjustbox}{width=\textwidth}
\begin{tabular}{lcccccccccc}
\toprule
 &
\makecell{Read-\\ability} &
\makecell{Reasoning\\Clarity} &
\makecell{Situational\\Fit} &
\makecell{Risk\\Ack.} &
\makecell{Fin. Harm\\Risk$^\dagger$} &
\makecell{Risk of Being\\Misled$^\dagger$} &
\makecell{Overall\\Quality} &
\makecell{Source\\Knowledge} &
\makecell{Trust\\Intention} &
\makecell{Reliance\\Intention} \\
\midrule

\multicolumn{11}{l}{\textit{Advice-style effects in the unlabeled arm}} \\
Expert vs. AI
& \sigpos{+0.383***}{0.43}
& \sigposlight{+0.257*}{0.30}
& \sigpos{+0.496***}{0.60}
& \nspos{+0.010}{0.01}
& \signeglight{-0.340*}{-0.27}
& \signeglight{-0.436*}{-0.32}
& \sigpos{+0.338**}{0.35}
& \nspos{+0.154}{0.17}
& \sigposlight{+0.283*}{0.29}
& \sigposlight{+0.328*}{0.33} \\[0.25em]

OC vs. AI
& \sigpos{+0.379**}{0.43}
& \nspos{+0.111}{0.13}
& \sigposlight{+0.232*}{0.28}
& \nspos{+0.087}{0.08}
& \nsneg{-0.113}{-0.09}
& \signeglight{-0.359*}{-0.27}
& \nspos{+0.205}{0.21}
& \signeg{-0.315**}{-0.35}
& \nspos{+0.097}{0.10}
& \nspos{+0.071}{0.07} \\[0.35em]

\midrule
\multicolumn{11}{l}{\textit{Label-arm effects for AI advice}} \\
Labeled vs. Unlabeled
& \nspos{+0.102}{0.12}
& \nspos{+0.076}{0.09}
& \sigposlight{+0.328*}{0.40}
& \nspos{+0.106}{0.10}
& \nsneg{-0.216}{-0.17}
& \nsneg{-0.107}{-0.08}
& \nspos{+0.301}{0.31}
& \nspos{+0.037}{0.04}
& \nspos{+0.142}{0.14}
& \nspos{+0.229}{0.23} \\[0.25em]

Mislabeled vs. Unlabeled
& \nspos{+0.058}{0.07}
& \nspos{+0.073}{0.09}
& \sigpos{+0.352**}{0.42}
& \nspos{+0.127}{0.12}
& \nsneg{-0.338}{-0.27}
& \nsneg{-0.322}{-0.24}
& \sigposlight{+0.402*}{0.42}
& \nspos{+0.065}{0.07}
& \nspos{+0.280}{0.28}
& \nspos{+0.295}{0.29} \\[0.35em]

\midrule
\multicolumn{11}{l}{\textit{Advice style $\times$  label arm interactions}} \\
Labeled $\times$ Expert
& \nspos{+0.096}{0.11}
& \nspos{+0.074}{0.09}
& \nsneg{-0.221}{-0.27}
& \nsneg{-0.146}{-0.14}
& \nsneg{-0.031}{-0.03}
& \nsneg{-0.061}{-0.05}
& \nspos{+0.056}{0.06}
& \nspos{+0.171}{0.19}
& \nspos{+0.055}{0.06}
& \nsneg{-0.021}{-0.02} \\[0.25em]

Labeled $\times$ OC
& \nspos{+0.160}{0.18}
& \nspos{+0.055}{0.06}
& \nsneg{-0.114}{-0.14}
& \nsneg{-0.362}{-0.35}
& \nspos{+0.194}{0.16}
& \nsneg{-0.042}{-0.03}
& \nsneg{-0.216}{-0.22}
& \nspos{+0.198}{0.22}
& \nspos{+0.038}{0.04}
& \nspos{+0.042}{0.04} \\[0.25em]

Mislabeled $\times$ Expert
& \nsneg{-0.018}{-0.02}
& \nsneg{-0.083}{-0.10}
& \signeglight{-0.300*}{-0.36}
& \nsneg{-0.205}{-0.20}
& \nspos{+0.051}{0.04}
& \nspos{+0.162}{0.12}
& \nsneg{-0.282}{-0.29}
& \nsneg{-0.091}{-0.10}
& \nsneg{-0.191}{-0.19}
& \nsneg{-0.266}{-0.27} \\[0.25em]

Mislabeled $\times$ OC
& \nspos{+0.092}{0.11}
& \nspos{+0.027}{0.03}
& \nsneg{-0.224}{-0.27}
& \signeg{-0.430*}{-0.41}
& \nsneg{-0.016}{-0.01}
& \nspos{+0.093}{0.07}
& \nsneg{-0.348}{-0.36}
& \nspos{+0.218}{0.24}
& \nsneg{-0.194}{-0.20}
& \nsneg{-0.067}{-0.07} \\[0.35em]

\midrule
$R^2_m$
& 0.211 & 0.156 & 0.123 & 0.100 & 0.107 & 0.090 & 0.117 & 0.075 & 0.129 & 0.162 \\[0.15em]

$R^2_c$
& 0.399 & 0.375 & 0.384 & 0.357 & 0.463 & 0.406 & 0.395 & 0.319 & 0.371 & 0.379 \\

\bottomrule
\end{tabular}
\end{adjustbox}

\vspace{0.5em}
\begin{minipage}{0.98\textwidth}
\footnotesize
\textit{Note.} This table reports unstandardized mixed-effects regression coefficients $\beta$ with standardized effect sizes $d$ in parentheses.
Models estimate advice style, label arm, and their interaction, with scenario fixed effects, scenario familiarity as a covariate, and participant-level random intercepts.
The reference condition is AI advice in the unlabeled arm. Thus, Expert vs. AI and OC vs. AI estimate advice style differences in the unlabeled arm; Labeled and Mislabeled coefficients estimate label-arm effects for AI-style advice; interaction terms indicate whether advice style differences change under labeled or mislabeled conditions relative to the unlabeled arm.
$N=285$ participants and $1{,}140$ vignette-level observations.
$R^2_m$ and $R^2_c$ denote marginal and conditional Nakagawa-style $R^2$, respectively.
$^\dagger$ Lower scores indicate more favorable evaluations for these outcomes.
Positive coefficients indicate higher ratings for the first-named condition; for daggered outcomes, negative coefficients indicate more favorable ratings for the first-named condition.
$^{*}p<.05$, $^{**}p<.01$, $^{***}p<.001$.
\end{minipage}
\end{table*}

\begin{figure*}[p]
    \centering
    \makebox[\textwidth][c]{%
        \includegraphics[
            page=1,
            width=1.06\textwidth,
            height=0.88\textheight,
            keepaspectratio,
            trim=10mm 16mm 8mm 14mm,
            clip
        ]{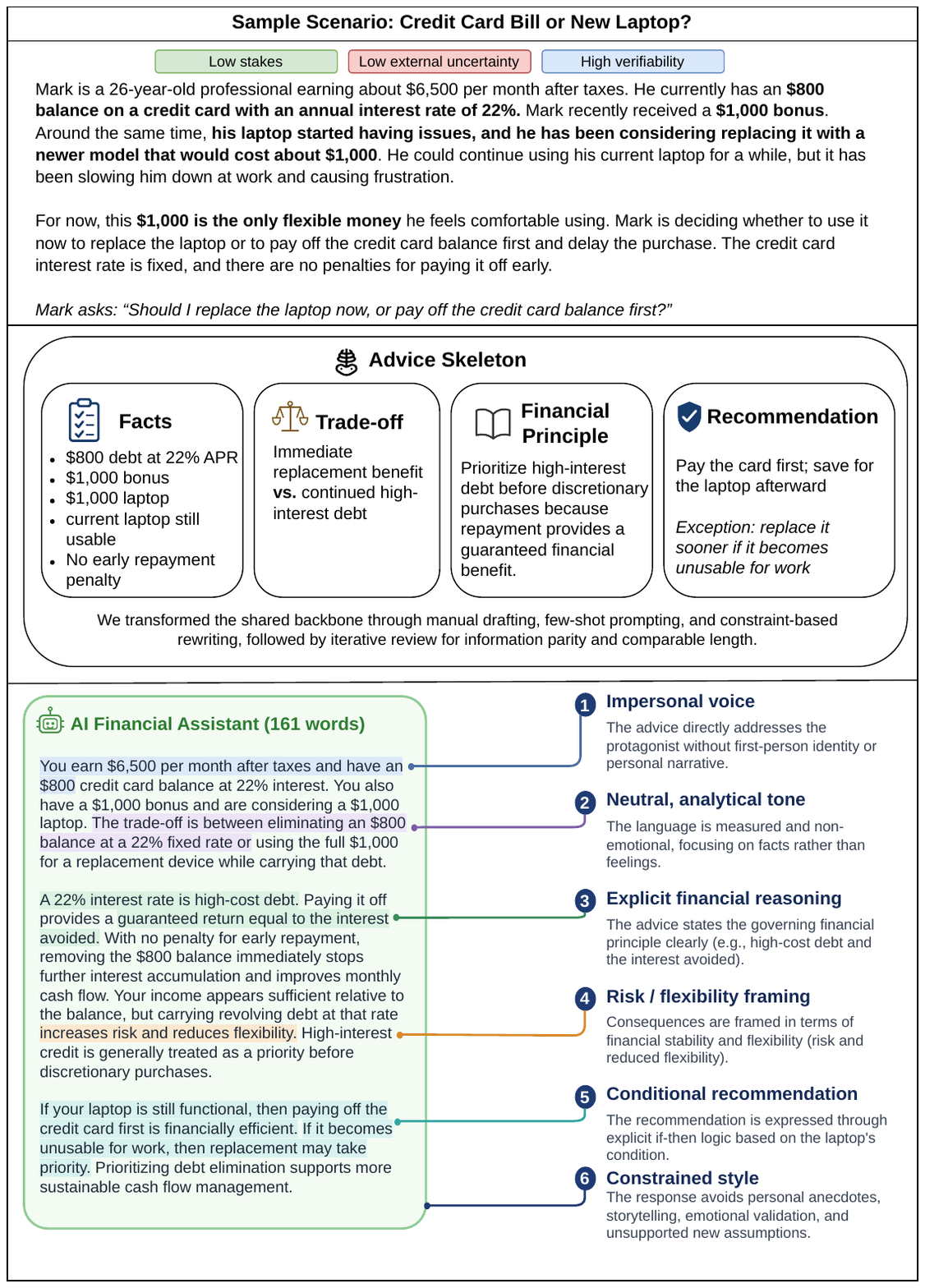}%
    }
    \Description{Page one of an annotated stimulus example. The credit-card-versus-laptop scenario is followed by a shared advice skeleton listing the facts, trade-off, financial principle, recommendation, and exception. A complete AI Financial Assistant response is highlighted and linked to numbered annotations identifying its impersonal voice, neutral analytical tone, explicit reasoning, risk framing, conditional recommendation, and constrained style.}

    \caption{
    \textbf{Controlled construction of source-specific advice styles.}
    A shared decision skeleton preserves the financial facts, governing
    principle, trade-off, recommendation direction, and boundary condition
    across all three responses. The same substantive backbone is rendered
    using the source-specific communication characteristics of an
    \emph{AI Financial Assistant}, a \emph{Certified Financial Planner}, and
    an \emph{Online Community Forum}. Highlighted passages and numbered
    annotations identify the voice, tone, framing, reasoning presentation,
    and discourse structure operationalized in each condition. Page one
    presents the financial scenario, common advice skeleton, and annotated
    AI Financial Assistant response.
    }
    \label{fig:annotated_advice_styles}
\end{figure*}

\begin{figure*}[p]
    \ContinuedFloat
    \centering
    \makebox[\textwidth][c]{%
        \includegraphics[
            page=2,
            width=1.06\textwidth,
            height=0.93\textheight,
            keepaspectratio,
            trim=10mm 30mm 10mm 7mm,
            clip
        ]{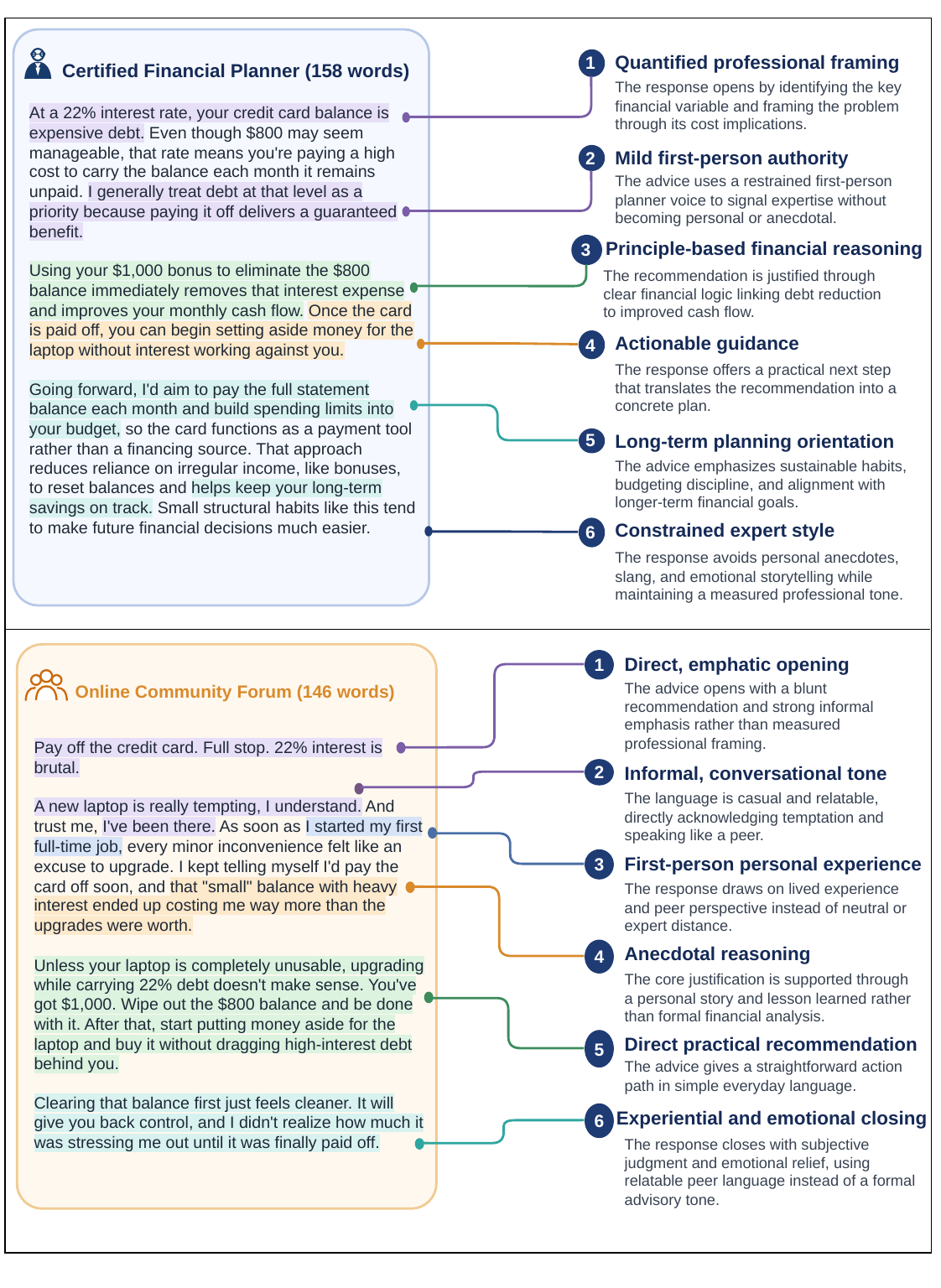}%
    }
    \Description{Page two of the annotated stimulus example. Complete Certified Financial Planner and Online Community Forum responses are highlighted and linked to numbered annotations. The planner response illustrates quantified professional framing, restrained first-person authority, principle-based reasoning, actionable guidance, and long-term planning, while the community response illustrates an emphatic opening, conversational tone, personal experience, anecdotal reasoning, direct practical advice, and an emotional closing.}

    \caption[]{
    \textbf{Controlled construction of source-specific advice styles
    (continued).}
    Annotated Certified Financial Planner and Online Community Forum
    responses generated from the same substantive advice skeleton.
    }
\end{figure*}


\end{document}

\endinput